# Optical fiber modal noise in the 0.8 to 1.5 micron region and implications for near infrared precision radial velocity measurements

Keegan S. McCoy[a], Lawrence Ramsey*[a,b], Suvrath Mahadevan[a,b], Samuel Halverson[a], Stephen L. Redman[c].

[a]Department of Astronomy & Astrophysics, The Pennsylvania State University, 525 Davey Laboratory, University Park, 16802, USA;
[b]Center for Exoplanets & Habitable Worlds, The Pennsylvania State University, University Park, PA 16802;
[c]Atomic Physics Division, National Institute of Standards and Technology, Gaithersburg, MD 20899, USA;

**ABSTRACT**

Modal noise in fibers has been shown to limit the signal-to-noise ratio achievable in fiber-coupled, high-resolution spectrographs if it is not mitigated via modal scrambling techniques. Modal noise become significantly more important as the wavelength increases and presents a risk to the new generation of near-infrared precision radial spectrographs under construction or being proposed to search for planets around cool M-dwarf stars, which emit most of their light in the NIR. We present experimental results of tests at Penn State University characterizing modal noise in the far visible out to 1.5 microns and the degree of modal scrambling we obtained using mechanical scramblers. These efforts are part of a risk mitigation effort for the Habitable Zone Planet Finder spectrograph currently under development at Penn State University.

## 1. INTRODUCTION

Baudrand, J. and G. Walker[1] first called attention to the importance of modal noise in fiber coupled spectrographs and suggested mechanical agitation of the fiber as the cure. The number of excited modes in a uniformly-illuminated fiber is given by:

$$M = \frac{1}{2}\left(\frac{\pi \cdot d \cdot NA}{\lambda}\right)^2 \qquad (1)$$

where $d$ is the fiber core diameter, $\lambda$ is the light's wavelength, and NA is the fiber's numerical aperture. Figure 1 displays the number of modes per wavelength in the optical and NIR for an optical fiber with a core diameter of 200 μm and NA = 0.22. In the visible the number of modes is large but substantially decreases as $\lambda^{-2}$ in the 1 to 1.7 micron region where the Habitable Zone Planet Finder (HPF) currently under development at Penn State will operate to measure precision radial velocities on M dwarf stars[2]. The fewer modes in the near infrared (NIR) make fiber coupled instruments operating there more susceptible to uncertainty in the instrument profile induced by modal noise. The effect of modal noise is primarily to introduce uncertainty in the position of the instrument profile and is distinct from the effect of incomplete scrambling[3] which can cause a shift in the instrument profile. Indeed, Ycas et al.[4] report that modal noise limited the radial velocity precision attainable with a frequency comb in the H band spectral region. To explore modal noise mitigation at a variety of wavelengths in the NIR and its mitigation we developed an experimental set up that allowed for rapid data acquisition and analysis.

*lwr@psu.edu; phone 1 814 863 5573; astro.psu.edu

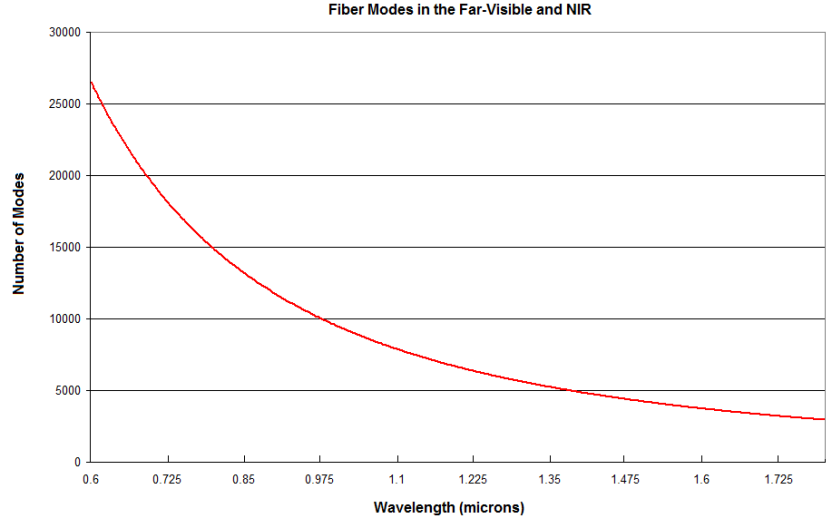

Figure 1. Number of modes in the far-visible and NIR for a multimode optical fiber (200 μm diameter core and 0.22 NA).

Coherent light entering a fiber within a particular angular range defined by the input focal ratio will excite a unique set of modes and produce a distinctive modal power distribution (MPD) at the fiber output. The MPD presents itself as a speckle pattern in the fiber's far-field, due to the interference of the electric-field patterns of the individual fiber modes[5]. A change in the input illumination of a fiber alters the MPD, and therefore, the far-field speckle pattern. Modal noise, the time variation of the MPD between successive exposures, is of particular concern when measuring precise RVs, since variation of the MPD will change the instrument profile (IP). Rawson et al.[6] noted three conditions necessary for modal noise: a sufficiently narrow source spectrum, spatial filtering at the output plane, and either a source wavelength shift, fiber movement, or both. Astronomical RV spectroscopy meets all three of these conditions, since each pixel in a high-resolution spectrograph's detector subtends a very narrow spectral width, the spectral slit and/or optics inevitably vignettes the spectrograph's input fiber, and the fiber link moves with the telescope[7]. To elaborate on the first criterion, modal noise is only a major concern for coherent light. When observing light of a broad bandwidth, the speckle contrast is dramatically reduced due to the much larger number of excited modes[8]. However, for very high-resolution spectrographs, the light at the detector becomes effectively coherent to the point that the SNR is no longer photon-limited, but modal noise-limited[9]; i.e. proportional to $M^{1/2}$.

## 2. EXPERIMENTAL APPROACH

### 2.1 Test Set Up

Our proxy for modal noise in the final spectrum is the far-field speckle pattern from a test fiber illuminated by a narrow band-source, e.g. a laser or laser diode. This has the advantage that visual inspection of the image allows a quick assessment of mitigation techniques and using straight forward image analysis techniques allow straight forward quantification. The perfect mitigation technique would change the speckle pattern during an exposure to a uniform disk. We obtain images of a speckle pattern (Figure 2) using either a Finger Lakes Instrumentation (FLI) deep depletion CCD or a Xenics NIR camera. A laser or laser diode is optically coupled to the 0.22 NA, 30 meter long Polymicro multimode fiber with a 200 μm diameter silica core, 20 μm thick cladding, and 40 μm thick polyimide buffer. As the sources vary greatly in brightness (table1) we use combinations of neutral density filters to avoid saturating the detectors. These sometimes stacked filters do lead to some fringing evident in some of the far field images (e.g. figure 3 left). The multimode fiber is terminated with FC connectors on both ends and the output end plugs into a 10 mm off axis mirror collimator. A weak lens adjusts this beam size slightly to nearly fill the detectors which have significantly different sizes. In the last meter or two before the collimator we apply agitation described below. We used the same approach at multiple wavelengths. Table 1 gives the sources, bandwidth and effective resolving power as well as output power.

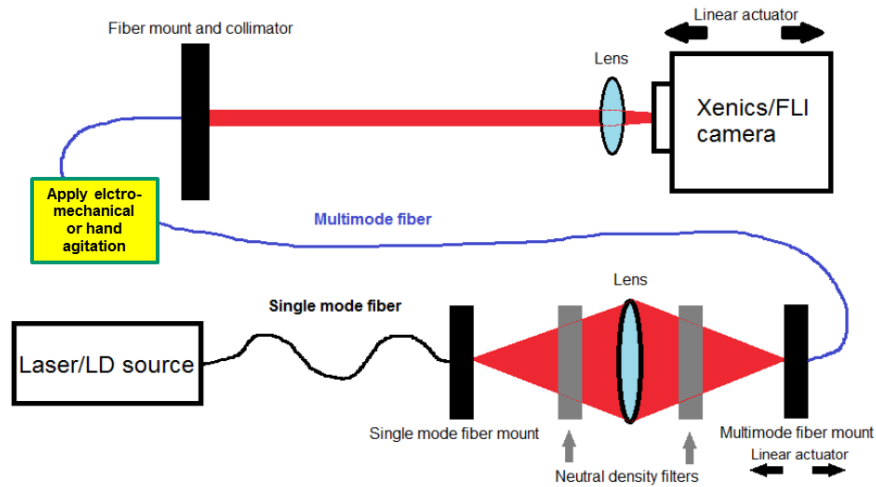

Figure 2. Schematic of optical coupling setup within the overall test configuration.

## 2.2 Coherent Light Sources

Modal noise was tested using 635 nm, 830 nm, 1310 nm, and 1550 nm Thor Labs laser diodes, a 632.8 nm Metrologic ML 810 HeNe laser (1.0 mW), and a 1550 nm Acronym Fiber Optics, Inc. (Item # DFB-C10) narrow-band, distributed feedback (DFB) laser. Table 1 contains the laser and laser diode specifications. A Thors Labs ITC 4001 Laser Diode/Temperature Controller was used to set the laser diode currents and stabilize the diode temperatures at 23 °C. The lasers and laser diodes were optically coupled to a 0.22 NA, 30 meter long Polymicro multimode fiber with a 200 μm diameter silica core, 20 μm thick cladding, and 40 μm thick polyimide buffer.

Table 1. Laser and laser diode specification (LD = laser diode).

| Wavelength | Spectral Bandwidth | Resolving Power ($\lambda / \Delta\lambda$) | Output Power |
|---|---|---|---|
| **LD 635 nm** | 0.08 nm | 7,938 | 2.5 mW |
| **Laser 632.8 nm** | 0.01 nm | 63,280 | 0.8 mW |
| **LD 830 nm** | < 0.1 nm | 8,300 | 10.0 mW |
| **LD 1310 nm** | ~ 1 nm | 1,310 | 2.5 mW |
| **LD 1550 nm** | 1.5 nm | 1,033 | 1.5 mW |
| **Laser 1550 nm** | 0.08 pm | 19,375,000 | 20 mW |

## 2.3 Detectors

For the data beyond 1000 nm we used a Xenics Xeva-1.7-320 NIR camera, which has a 256 x320 pixel array with peak sensitivity from 900–1700 nm (22% QE at 635 nm and 84% QE at 1550 nm). In addition to the Xenics camera used for the NIR tests, a 1056 x 1027 pixel Finger Lakes Instrumentation (FLI) PL4710 deep depletion CCD camera with 300–1000 nm wavelength coverage was used to provide higher spatial resolution for the 632.8, 635 and 830 nm data. The FLI's E2V CCD has 13 μm pixels, compared to the 30 μm pixels of the Xenics camera's NIR array.

## 2.4 Experimental Procedure

Three individual sets of 10 frames in sequence were taken for each of the four laser diodes and two lasers. The first sets were taken with the multimode fiber static (referred to as the static test), the second sets with the multimode fiber vibrated at 60 Hz by a FMC Syntron PowerPulse electromagnetic paint mixer (mechanical vibration test), and the third sets with the multimode fiber agitated by hand at ~1–2 Hz with an amplitude of ~10–15 cm (hand agitation test). For the hand agitation tests, the multimode fiber was held with two hands approximately 25 cm apart. Each hand was alternatively moved up and down 180° out of phase at approximately the above frequency and amplitude. Individual test frames were taken with a 3 second integration time to average over a significant number of vibrational periods or hand movement and to provide a sufficient SNR.

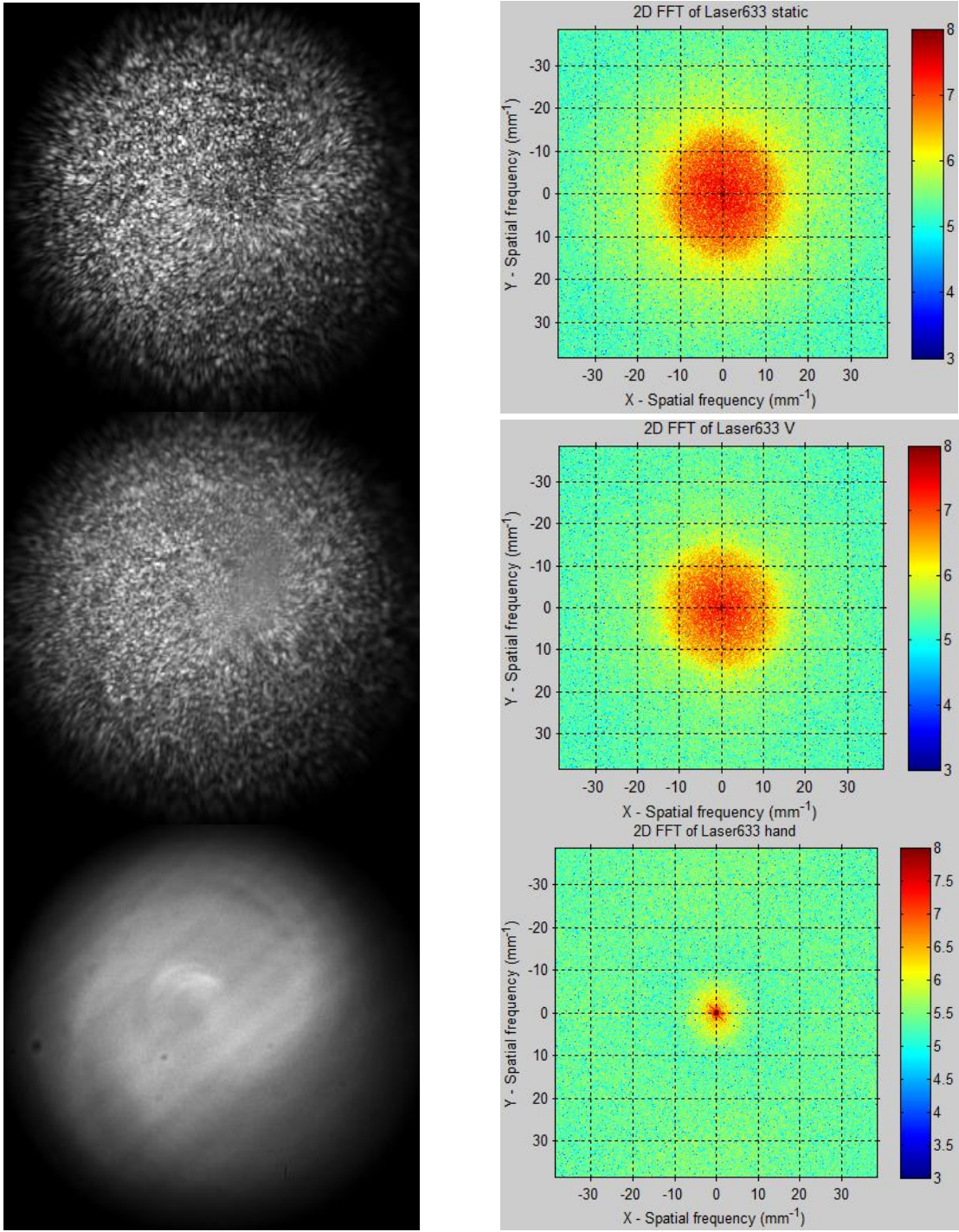

Figure 3. 632.8 nm HeNe laser tests (FLI camera) – (top left) static image, (top right) static image 2D FFT, (middle left) mechanical vibration image, (middle right) mechanical vibration image 2D FFT, (bottom left) hand agitation image, (bottom right) hand agitation image 2D FFT.

**2.5 Data Analysis**

The visual representations of these test results are helpful, but a spatial frequency analysis is required to fully characterize modal variations. The right hand side of Figure 3 above shows the 2D Fast Fourier Transforms (2D FFT) for the corresponding speckle images to the left . Figure 4 below displays the 2D rotationally-averaged spatial frequency power spectrum plots for all test cases in Figure 3 test cases, with normalized power and spatial frequency axes. The Xenics NIR camera has 0.03 mm pixels and so spatial frequencies are given in cycles/mm, or simply $mm^{-1}$, and plotted out to the Nyquist frequency (1 cycle / 2 pixels = 16.67 $mm^{-1}$). The smaller 13 μm pixels of the FLI camera provide 2.3 times the linear spatial resolution of the Xenics camera's 30 μm pixels and yields a Nyquist frequency of 1 cycle / 2 pixels = 38.46 $mm^{-1}$) . Since fibers are vibrated to scramble the power in the various excited modes and homogenize the far-field speckle pattern, a perfectly scrambled multimode fiber with an idealized top-hat far-field illumination pattern would only consist of the a narrow Sync function with most of the power at frequencies less than 1/W $mm^{-1}$. W is the width of the far field pattern which is ~ 5-10 mm in our data.

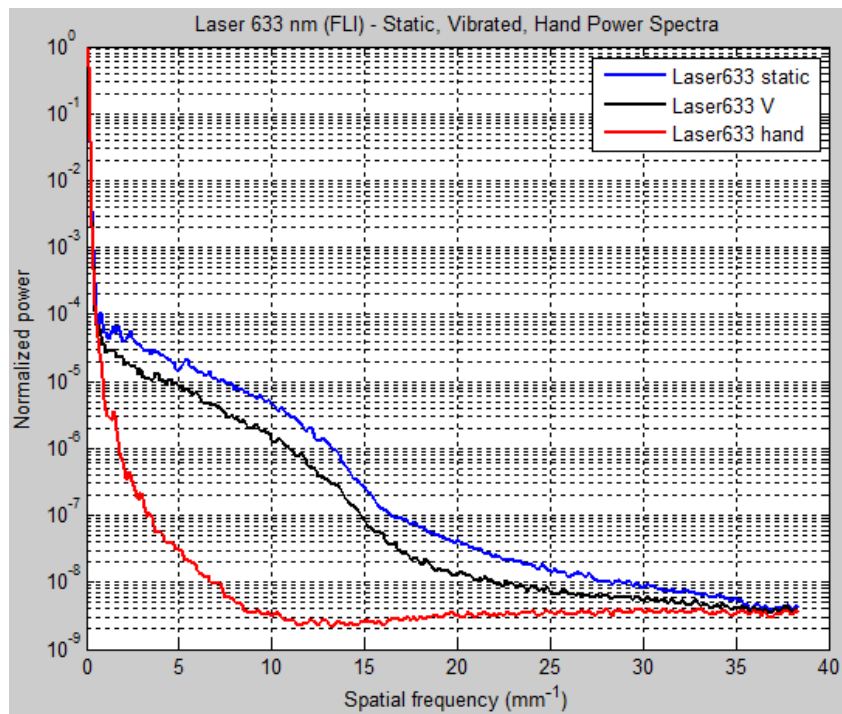

Figure 4. 632.8 nm HeNe laser tests power spectra (FLI camera) – (blue line) static image power spectrum, (black line) mechanical vibration image power spectrum, (red line) hand agitation image power spectrum..

## 3. RESULTS

In Figure 5 we show the circularly averaged power spectra for the four laser diodes for the static, electromechanically vibrated (V) and hand agitated case. The top two are for the 635nm diode (left) and 835 nm diode (right) and were taken with the FLI CCD camera. It is clear that the higher spatial resolution of this camera well resolves the speckles even in the static case, as the power spectrum flattens to white noise in both cases at the higher frequencies. The higher wavelength bandwidth of these diodes is apparent when compared to the HeNe laser in Figure 4 as there is little to no power beyond ~17 $mm^{-1}$. The consistent result for all four diodes wavelengths is that rapid low amplitude (few mm) vibration of the electromechanical 60Hz vibrator is not reducing modal noise at a level near that of larger amplitude hand agitation.

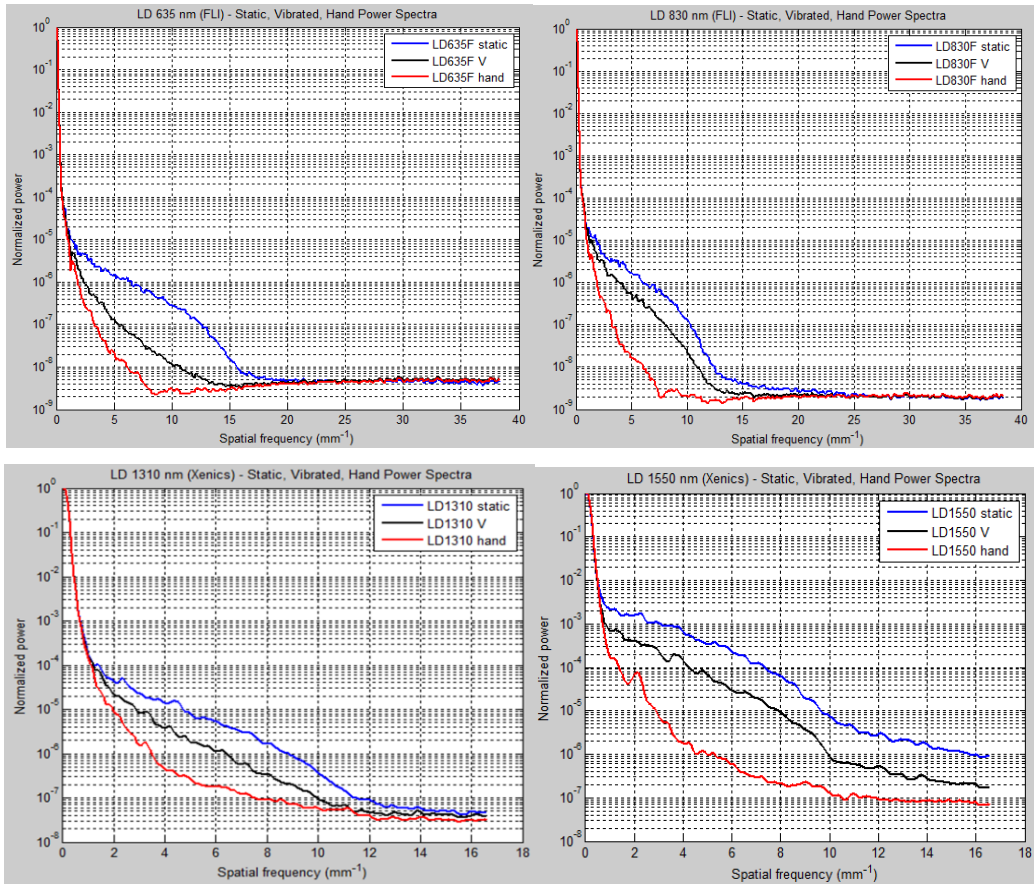

Figure 5. Power spectra for the laser diodes in Table 1 at 635, 830, 1310 and 1550 nm.

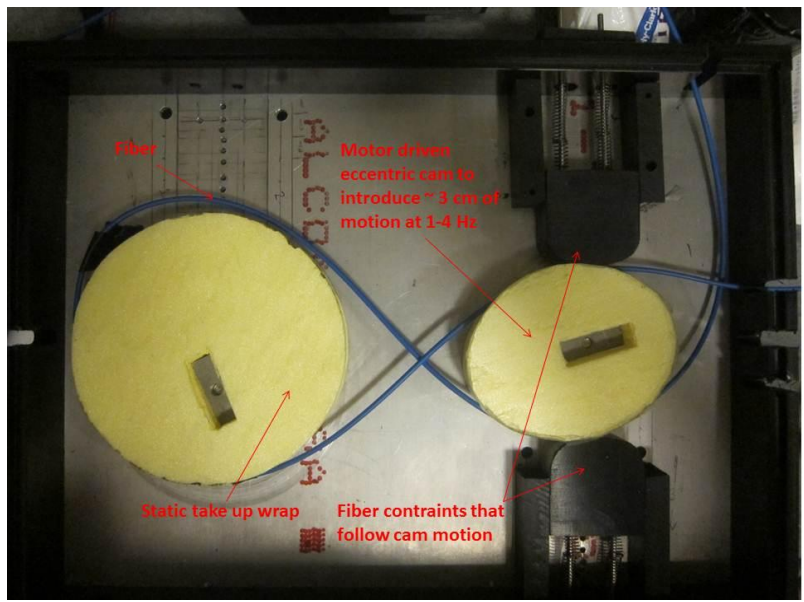

Figure 6. Prototype large(r) amplitude mechanical agitator.

The fact that large amplitude hand agitation was substantially more effective in reducing modal noise led us to develop the lower frequency but larger amplitude mechanical agitator shown in Figure 6. In this test device the test fiber is wound around a static circular take-up spool (on the left in Figure 6) and moved up and down by an eccentric cam (on the left in Figure 6) by a few cm at a frequency of a 2-3 Hz by a DC motor. The fiber is kept in a groove on the eccentric cam by two spring loaded capstans. Using the narrow band 1550 nm laser as the most severe test we achieved the results in Figure 7. This figure shows the mechanical agitator in Figure 6 is almost as good as the hand agitation. Note the fringing apparent in the images is likely due to stacked neutral density filters between the Laser/Laser Diode source and the 200 micron core test fiber as noted in section 2.1. This fringing patter is also clear in the power spectrum and in principle could be filtered out for the modal noise test measurements.

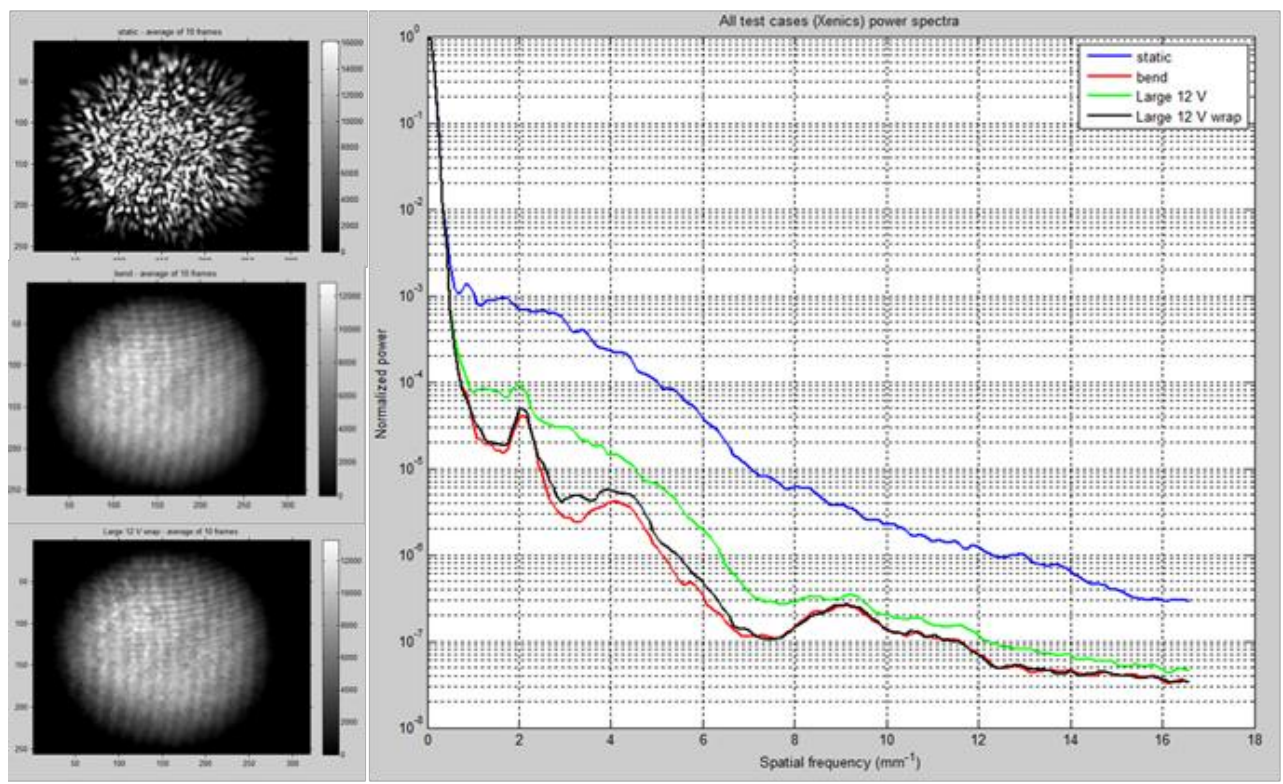

Figure 7. On the left top to bottom we show the static, hand agitated as described (called bend here) and mechanically agitated speckle pattern averaged over 10 frames. On the right is the power spectrum of these cases including an intermediate case (green) where the fiber was not wrapped around the take up spool.

## 4. CONCLUSIONS AND NEXT STEPS

Modal noise, which is significantly worse in the NIR, poses a substantial risk to achieving m/sec precision radial velocities in NIR instruments and must be mitigated. If left uncorrected it will severely impact the performance of the next generation of stable, fiber-fed NIR spectrographs like HPF, CARMENES, and SPiRoU. Using speckle pattern as a proxy for modal noise we have developed a mechanical agitation approach that will mitigate modal noise to a very high level on Habitable Zone Planet Finder spectrograph being developed at Penn State. One further test that must be done is to look at the effect of macro mechanical agitation on the fiber lifetime. We are re-designing the agitator in Figure 6 to more closely mimic the hand motion described in section 2.4 which appears to be more effective than the rotating cam. We are also planning to use our Pathfinder[10] test bed instrument to calibrate the modal noise reduction we see in the power spectra of the speckle pattern with the stability of the instrument profile.

## ACKNOWLEDGEMENTS

Dan Anderson, Kyle McCausand and Arpita Roy contributed to the tests with the mechanical agitator. This work was partially supported by the Center for Exoplanets and Habitable Worlds, which is supported by the Pennsylvania State University, the Eberly College of Science, and the Pennsylvania Space Grant Consortium. We acknowledge support from NSF grant AST-1006676, AST-1126413, the NASA Astrobiology Institute (NAI), and PSARC. This research was performed while SLR held a National Research Council Research Associateship Award at NIST. Much of this work constitutes Mr. Keegan McCoy's MS thesis in Electrical Engineering and as a student he was supported this last year by the United States Air Force.